\documentclass[prl,twocolumn,showpacs,superscriptaddress]{revtex4-1}
\usepackage{graphicx,amssymb,amsmath,color}
\begin{document}
\definecolor{darkgreen}{rgb}{0,0.5,0}
\newcommand{\be}{\begin{equation}}
\newcommand{\ee}{\end{equation}}
\newcommand{\jav}[1]{{#1}}

\title{Absence of orthogonality catastrophe after a spatially\\ inhomogeneous interaction quench in Luttinger liquids}

\author{Bal\'azs D\'ora}
\email{dora@eik.bme.hu}
\affiliation{Department of Physics and BME-MTA Exotic  Quantum  Phases Research Group, Budapest University of Technology and
  Economics, 1521 Budapest, Hungary}
\author{Frank Pollmann}
\affiliation{Max-Planck-Institut f\"ur Physik komplexer Systeme, 01187 Dresden, Germany}

\date{\today}

\begin{abstract}
We investigate the Loschmidt echo, the overlap of the initial and final wavefunctions of Luttinger liquids after a spatially inhomogeneous interaction quench. 
In studying the Luttinger model, we obtain an analytic solution of the bosonic Bogoliubov-de Gennes equations after quenching the interactions within a finite spatial region. 
As opposed to the power law temporal decay following a potential quench, the interaction quench in the Luttinger model leads to a finite, 
hardly time dependent overlap, therefore no orthogonality catastrophe occurs.
The steady state value of the Loschmidt echo after a sudden inhomogeneous quench is the square of the respective adiabatic overlaps.
Our results are checked and validated numerically on the XXZ Heisenberg chain.
\end{abstract}


\maketitle

\emph{Introduction.} The sensitivity of a quantum time evolution to perturbations is a central problem in many distinct areas of physics.
Starting probably from a discussion between J. Loschmidt and
L. Boltzmann, the effect of time forward and reversed processes has always enjoyed the attention of prominent researchers.
The revival of interest towards non-equilibrium time evolution has been triggered by recent experiments in cold atomic gases and questions such as thermalization and equilibration  \cite{kinoshita}, defect 
production during passage through a  quantum critical point  \cite{polkovnikovrmp, dziarmagareview}, quantum work fluctuation relations  \cite{rmptalkner} 
etc.   call for 
developments from both the experimental and theoretical side.

Non-equilibrium states can be reached in many different ways. While the condensed matter thinking typically implies the application of strong electric and magnetic fields  \cite{aokibook},
cold atomic systems offer the possibility to change interactions by tuning to or away from a Feshbach resonance or by altering the lattice parameters  \cite{BlochDalibardZwerger_RMP08}.
From the latter class, quantum quenching the interaction strength has been studied in a variety of systems \cite{polkovnikovrmp, dziarmagareview}. The common feature in these approaches
is an abrupt change of a spatially homogeneous interaction parameter. However, the consideration  of spatially inhomogeneous quenches can be equally exciting \cite{sotiriadis,foster,lancaster,ramsdziarmaga}.
The famous example is the x-ray edge problem \cite{nersesyan}, where a spatially local (therefore highly inhomogeneous in $k$ space) potential scatterer is switched on abruptly, realizing the time-dependent 
version of Anderson's orthogonality catastrophe (OC).

Inspired by these, 
we focus on the interplay of spatial inhomogeneity and strong correlations on the non-equilibrium dynamics by studying an inhomogeneous interaction quench in a Luttinger liquid (LL).
The prototypical dynamical quantity, central to  x-ray edge physics, is the Loschmidt echo (LE) \cite{peres}, which is  the overlap of two wave functions, $|\Psi_0(t)\rangle$ and 
$|\Psi(t)\rangle$, evolved from the same
initial state,  but with different Hamiltonians, $H_0$ and $H$,
\begin{equation}
\mathcal{L}(t)\equiv \left|\langle\Psi_0(t)|\Psi(t)\rangle\right|^2.
\end{equation}

\jav{Inhomogeneous quantum quenches have been well studied in various systems\cite{affleck,schiro,smacchia,gambassi,rossini,nersesyan}.
For relevant perturbations, the LE decays in a power law fashion with universal exponent, while 
for a marginal defect, the exponent depends on the final strength of the perturbation\cite{smacchia}.
After an inhomogeneous quench over a finite spatial region $l$, excitations are mostly produced within the quenched region. 
A potential quench  of strength $h$ within a region $l$ keeps the LE  close to 1} for times $t\ll l/\mathsf{v}_F$ ($\mathsf{v}_F$ the maximal propagation velocity) 
as the quasiparticles in the quenched region feel a chemical potential shift, which does not destroy the coherence.
For long times $t\gg l/\mathsf{v}_F$, the created excitations have left
the quenched region, and the overlap decays in a power law fashion. The exponent is determined by the phase shift $\sim (lh/\mathsf{v}_F)^2$ without additional interactions, when the potential is marginal.
\jav{In the presence of repulsive interactions, i.e. in a Luttinger liquid, a local potential is a relevant perturbation, yielding a universal decay exponent\cite{affleck,smacchia}.}

For  an inhomogeneous \emph{interaction} quench, the focus of this work, the LE in the short time regime is expected to scale
 similarly to that after a homogeneous quench, after replacing the system size by $l$. The long time response of the system is, however, not obvious at all.
\jav{Based on the x-ray edge problem, one would expect the LE to vanish with $t$, since local backscattering terms are inevitably induced at the boundary of the interacting region\cite{sedlmayr}.}
As opposed to this, we show that no OC occurs for 
an inhomogeneous interaction quench \jav{in one spatial dimension} and the LE in the $l\gtrless \mathsf{v}_F t$ regions remains finite.
Our analytical calculations are checked and confirmed numerically using time dependent
density matrix renormalization group (tDMRG)  \cite{Vidal2004, Daley2004,FeiguinWhite2005,Schollwock2011} based methods on the XXZ Heisenberg model,
which contains local backscattering terms at the boundary of the interacting region.

Our work is interesting and relevant not only for its condensed matter and cold atomic aspect, but bears importance in nuclear physics as well where
the overlap of (fermionic) Hartree-Fock-Bogoliubov wavefunctions is essential in determining the properties of nuclear states \cite{robledo,onishi} as 
it accounts for deformation and pairing.

\begin{figure}[h!]
\includegraphics[width=7cm]{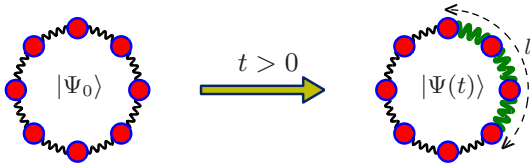}
\caption{The low energy dynamics of the XXZ Heisenberg chain is faithfully represented by the Luttinger model.
An inhomogeneous interaction quench in the Luttinger model is equivalent to the continuum limit of a harmonic chain, sketched in the figure,
where the spring constants (and/or masses) of a region of size $l$ are altered abruptly at $t=0$.}
\label{harmonicchain}
\end{figure}

\emph{Inhomogeneous Luttinger liquids.} We investigate the spatially inhomogeneous and time dependent XXZ Heisenberg model, which reads as
\begin{gather}
H=\sum_{n=-L/2}^{L/2-1}J\left(S_n^xS_{n+1}^x+S_n^yS_{n+1}^y\right)+J_z(t)\sum_{n=-l/2}^{l/2-1} S_n^zS_{n+1}^z
\label{xxz}
\end{gather}
where $n$ indexes the lattice sites, and $J>0$ is the antiferromagnetic exchange
interaction.  We are going to manipulate $J_z<J$ via a sudden quench as $J_z(t)=J_z\Theta(t)$, with $\Theta(t)$ the Heaviside function.  
The spatial width of the interaction is $l$, and smoothing the envelope function does not alter our results.
The effective low-energy dynamics of the Heisenberg model is described by an ``elastic string'' \cite{giamarchi,nersesyan}, whose first quantized form is
\begin{gather}
H_0=\frac{\mathsf{v}_F}{2\pi}\int\limits_{-L/2}^{L/2} dx\left[\left(\pi\Pi(x)\right)^2+\left(\partial_x \phi(x)\right)^2\right],
\label{gswf}
\end{gather}
where $\psi$ and $\Pi$ are conjugate variables with $[\phi(x),\Pi(y)]=i\delta(y-x)$ and $L$ is the length of the system.
The inhomogeneous interaction quench of the Heisenberg model induces 
\begin{gather}
H_{\mathrm{\mathrm{int}}}(t>0)=\frac{gK_0}{2\pi}\int_{-l/2}^{l/2} dx \left(\partial_x\phi(x)\right)^2, 
\label{hint}
\end{gather}
where $K_0$ emanates from a homogeneous interaction in the initial state and $l=L$ represents a global quench.
 The LL parameter in the quenched region is $K=K_0/\sqrt{1+(gK_0/\mathsf{v}_F)}$, while the renormalized velocity is $\mathsf{v}=\mathsf{v}_FK_0/K$.
For the Heisenberg model, the LL parameter is $K=\pi/2[\pi-\arccos(J_z/J)]$ \cite{giamarchi}.
The effective Gauss model is sketched in Fig. \ref{harmonicchain}.
The equilibrium and transport properties of the spatially inhomogeneous LLs  have long been investigated \cite{safi,maslov,ponomarenko,thomale}.
{Note that a spatially abrupt interaction barrier in Eq. \eqref{xxz} induces also backscattering\cite{sedlmayr}, neglected in the above continuum description.
Surprisingly, the LE is not affected by these terms as demonstrated below by a careful comparison between bosonization and tDMRG.}

The time evolved wavefunction,  $\Psi(t)=U(t)\Psi_0$, where the time evolution is governed by Eq. \eqref{hint}
as
\begin{gather}
U(t)=\mathcal T\exp\left(-i\int_0^t H_{\mathrm{int}}(t')dt'\right).
\label{u(t)}
\end{gather}
This time evolution operator can be calculated using a linked cluster or cumulant expansion techniques \cite{mahan,nersesyan}, similarly to how the X-ray edge problem was approached.
However, for the present problem, it is more advantageous to generalize the results {of the homogeneous system\cite{doraLE,agarwal} for the inhomogeneous case.}

The total Hamiltonian is rewritten as
\begin{gather}
H(t>0)=\Phi^{\dagger}
\left(\begin{array}{cc}
\Omega & G \\
G & \Omega
\end{array}\right)
\Phi
\end{gather}
where $\Phi^\dagger=(b^\dagger_{k_1},b^\dagger_{k_2},\dots b^\dagger_{k_N},b_{k_1},b_{k_2},\dots b_{k_N})$, 
and $k_{1\dots N}$ are integer multiples of $2\pi/L$, excluding $k=0$. 
The symmetric blocks are defined as
\begin{gather}
\Omega(k,p)=\mathsf{v}_F|k|\delta_{k,p}+\frac{g K_0}{2L}\sqrt{|kp|}\tilde f(k-p,l),\\
G(k,p)=\frac{g K_0}{2L}\sqrt{|kp|}\tilde f(k+p,l),
\end{gather}
and $\tilde f(k,l)$ is the Fourier transform of the envelope function of the interaction, which is $\tilde f(k,l)=2\sin(kl/2)/k$ for an abruptly terminated interaction over $l$ [see Eq. \eqref{xxz}]. 
For $t<0$, $g=0$.
The matrix structure of $\Omega$ and $G$ naturally favours a sharp momentum cutoff $q_c$.
The time evolution of the bosons is determined from the Heisenberg equation of motion, whose solution is written in the concise form
as $\Phi(t)=U(t)\Phi(0)$, where the time evolution operator defines the  generalized Bogoliubov matrices, $u(t)$ and $v(t)$, similarly to the homogeneous case \cite{doraLE} as
\begin{gather}
U(t)=\exp\left[-it\left(\begin{array}{cc}
\Omega & G \\
-G & -\Omega
\end{array}\right)\right]\equiv
\left(\begin{array}{cc}
u & v\\
v^* & u^*
\end{array}\right),
\end{gather}
and the LE is expressed in terms of the regularized Fredholm determinant \cite{berezin}
\begin{gather}
\mathcal{L}(t)=1/\sqrt{\det(1+v^\dagger v)},
\label{leexp}
\end{gather}
and  $v^\dagger=-v$. The fermionic version of Eq.~\eqref{leexp} in equilibrium is known as the Onishi formula \cite{onishi}.

Using Ref.  \cite{tegmark}, 
the generalized Bogoliubov coefficients are expressed in terms of the Hamiltonian matrix blocks as
\begin{gather}
v=\frac{\cos(t\tilde\omega)-\cos(t\tilde\omega)^T}{2}+i\frac{\Omega}{2}
\left(\frac{\sin(t\tilde\omega)}{\tilde\omega}-\left(\frac{\sin(t\tilde\omega)}{\tilde\omega}\right)^T\right)+\nonumber\\
+i\frac{G}{2}\left(\frac{\sin(t\tilde\omega)}{\tilde\omega}+\left(\frac{\sin(t\tilde\omega)}{\tilde\omega}\right)^T\right),
\label{vexact}
\end{gather}
where $\tilde\omega=\sqrt{(\Omega-G)(\Omega+G)}$, and $u$ is obtained similarly, though it is not needed for our purposes.
The commutator $[\tilde\omega,\tilde\omega^T]\neq 0$ for an inhomogeneous quench, and vanishes only for a homogeneous quench, when only the last term is present in $v$.
With Eq. \eqref{leexp} and \eqref{vexact},
the asymptotic long time limit of the overlap (taken after the thermodynamic limit is taken) is given by
\begin{gather}
\mathcal{L}_\infty\equiv \mathcal{L}(t\rightarrow\infty)=\det\left[\frac 12+\frac{\Omega}{4}\left(\tilde\omega^{-1}+\tilde\omega^{-1T}\right)+\nonumber\right.\\
\left.+
\frac{G}{4}\left(\tilde\omega^{-1}-\tilde\omega^{-1T}\right)\right]^{-1}.
\label{linfty}
\end{gather}
The long time limit of the sudden quench overlap is the square of the overlap of the adiabatic ground state wavefunctions, $\mathcal{L}_{\infty}=\mathcal{L}_{gs}^2$. This extends the previous results in  \cite{doraLE} 
to  the more general inhomogeneous quench case.
A finite value of $\mathcal{L}_\infty$ would indicate the absence of an OC for an inhomogeneous interaction quench.
The heart of the evaluation is the $v^\dagger v$ matrix, which is diagonal in the homogeneous case, yielding
the result of Ref.  \cite{doraLE}. 
In the following we will analyze various special cases of Eq.~(\ref{linfty}).

\begin{figure}[h!]
\includegraphics[width=7cm]{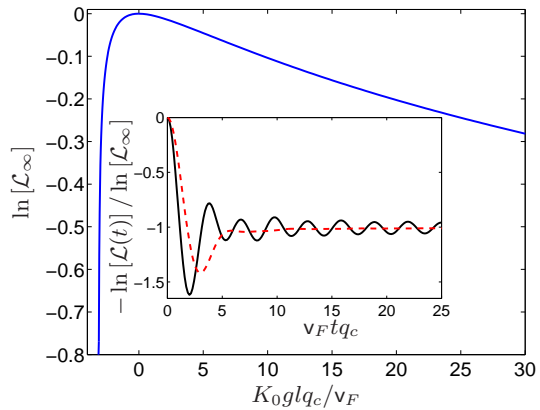}
\caption{The long time limit of the LE after a local interactions quench for $Lq_c=4000$ is 
evaluated for the Luttinger model from Eq. \eqref{leexp}. The inset depicts the time evolution for $K_0glq_c/\mathsf{v}_F=-1$ and 2 from Eq. \eqref{linfty}.}
\label{lelocal}
\end{figure}

\emph{Local quench region.} ($q_c l\ll 1$):
This limit is obtained by taking $l\rightarrow 0$ but keeping $gl$ fixed, therefore $\tilde f(k,l)=l$, and the LE depends only on the dimensionless number $K_0glq_c/\mathsf{v}_F$.
This local interaction quench limit does not apply directly to LLs, because many additional local terms, 
neglected in the Gauss model, can play an important role as is the case for the X-ray
edge problem \cite{nersesyan}. Nevertheless, it describes a variety of other problems as e.g. a single impurity immersed in a Bose gas \cite{zipkes,gaul}  or a 
molecular, localized defects in a quantum harmonic chain \cite{bruhl}. Moreover, 
it reveals the essential differences between a local \emph{interaction} and \emph{potential} quench
\footnote{ Eq. (\ref{u(t)}) can be evaluated using a cumulant expansion. For a local potential quench, its amounts
to calculate a Cauchy determinant \cite{yuval}, i.e. the determinant of a matrix with elements $a_{i,j}\sim 1/(i-j)$. For a local interaction quench, 
the evaluation of a generalized Cauchy determinant is required with elements $a_{i,j}\sim 1/(i-j)^2$, for which no useful results are available. We rely on the inhomogeneous Bogoliubov equations and 
Eq. (\ref{leexp}) instead.}.
The system remains stable for $K_0glq_c>-\pi \mathsf{v}_F$, otherwise one of the frequencies becomes imaginary signaling an instability. Using the mapping \cite{mahan} between quadratic bosonic models
and a system of coupled harmonic oscillators, the confining parabola of the oscillators flattens and becomes inverted upon decreasing $g$ further.
The short time decay features the universal time dependence   as
$\ln\mathcal{L}(t)\approx -t^2\textmd{Tr}\left(G^2\right)\sim-\left({{g}K_0q_c^2 t l}\right)^2,$
and 
the prefactor of the $t^2$ term is the variance of energy after the quench as expected on general grounds \cite{peres}.
In the long time limit (taking $L\rightarrow \infty$ first), the overlap tends to a finite, non-zero value (see Fig. \ref{lelocal}), in sharp contrast to the case of a potential quench,
signalling the absence of OC for the Luttinger model in the case of a local interaction quench. 

\emph{Quenching over a finite spatial region} ($q_c l\gg 1$).
In this limit, the LE picks up distinct contributions from small and large momentum states.
Due to the structure of the envelope function $\tilde f(k,l)$, a given momentum state $p$ can only be scattered within a $p\pm \pi/l$ momentum window,
which becomes rather narrow with increasing $l$. In the $q_c l\gg 1$ limit, for large momentum states with $|p|\gg\pi/l$, the momentum remains an almost conserved quantum number,
as its uncertainty, $\pi/l$ is much smaller than its typical value $p$. Averaging over states within the narrow momentum shell $\pm \pi/l$,
momentum conservation is regained in the $|p|\gg\pi/l$ region at the expense of enlarging the phase volume from $2\pi/L$ to $2\pi/l$.
Therefore, the corresponding LE after an inhomogeneous quench is identical to that after a global quench \cite{doraLE}, after replacing the 
minimum phase space volume, $2\pi/L$ by $2\pi/l$ \footnote{Changing the minimal momentum from 0 to $\pi/l$ gives an $O(1)$ correction, thus it is negligible}, \emph{regardless} to the explicit value of $t$.
In addition, the small momentum states states with $|p|\ll \pi/l$ are spatially extended  compared to $l$, therefore 
experience strong momentum scattering as if a local interaction quench has occurred, and the results of the previous section applies for these modes after replacing
the upper momentum cutoff, $q_c$ with $\pi/l$, yielding an $l$ independent dimensionless interaction strength as $K_0 glq_c/\mathsf{v}_F \rightarrow K_0g\pi/\mathsf{v}_F$. Since $l$ appeared only in this
combination for a local quench, the contribution from states $|p|\ll \pi/l$ to the LE is independent of $l$.
\begin{figure}[h!]
\includegraphics[width=7cm]{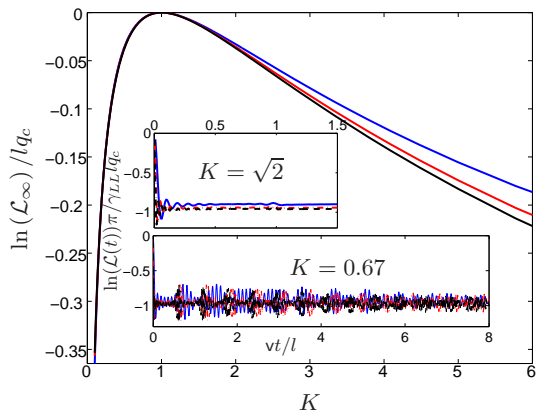}
\caption{The long time limit of the LE for the Luttinger model with $Lq_c=2000$ is shown for $lq_c=40$ (blue), 120 (red)  and 500 (black) evaluated from Eq. \eqref{linfty}, 
the latter is indistinguishable from that after a homogeneous quench.
The insets show the time dependence of the LE for the Luttinger model for various $K$'s and $lq_c=40$ (blue solid line), 80 (red dash-dotted line) and 120 (black dashed line) from Eq. \eqref{leexp}.}
\label{legauss}
\end{figure}
Consequently, the LE of an inhomogeneous quench, to leading order in $l$, is
\begin{gather}
\frac{\ln \mathcal{L}(t)}{l}=-\int\limits_{0}^{q_c} \frac{dq}{2\pi} \ln\left(1+\frac{\sin^2(\omega({q})t)}{4}\left(\frac{K}{K_0}-\frac{K_0}{K}\right)^2\right),
\label{LEinh}
\end{gather}
with $\omega(q)=\mathsf{v}|q|$.

In the long time limit with  $t\gg 1/\mathsf{v}q_c$ limit, this gives
\begin{gather}
\mathcal L_\infty(t\gg 1/\mathsf{v}q_c)=\exp\left(-\gamma_{LL}\frac{lq_c}{\pi}\right)
\label{leSQ}
\end{gather}
with
\begin{gather}
\gamma_{LL}=\ln\left(\frac 12+\frac 14\left(\frac{K}{K_0}+\frac{K_0}{K}\right) \right),
\end{gather}
which is related to the quantum geometric tensor of the SU(1,1) Lie group \cite{provost1980}.

The short time decay, calculated from Eq. \eqref{leexp},  features again the  universal time dependence  in the $t\ll 1/\mathsf{v}q_c$ limit as
$\ln\mathcal{L}(t)\approx -t^2\textmd{Tr}\left(G^2\right)=-c\left({{g}K_0q_ct}\right)^2 q_cl$,
with $c$ a non-universal constant. Using $gK_0=\mathsf{v}(K_0/K-K/K_0)$, this agrees with the short time expansion of Eq.~\eqref{LEinh}.

For intermediate times, the LE oscillates around its steady state value with a frequency increasing with $g$, 
and the oscillations are under/overdamped for $g\gtrless 0$, as shown in Fig. \ref{legauss}. 
The sharp oscillations arise mostly from the sharp cutoff scheme, a smoother (e.g. exponential) cutoff smoothens the  oscillations.
These are similar to the collapse and revival phenomenon of the LE in finite systems \cite{venuti1,peres}, although
some of the excitation, created during the inhomogeneous quench, can be transmitted to the unquenched region,
which then propagate freely and do not contribute to revivals any more as  these do not interfere with other excitations.
The steady state value of the LE agrees nicely with our analytical prediction in Eq. \eqref{leSQ}. For small $l$, slight deviations are visible with increasing $K$ ($K_0g\rightarrow -\mathsf{v}_F$) due to the contribution from
small momentum states mostly, which experience a local interaction quench.

\begin{figure}[h!]
\includegraphics[width=8cm]{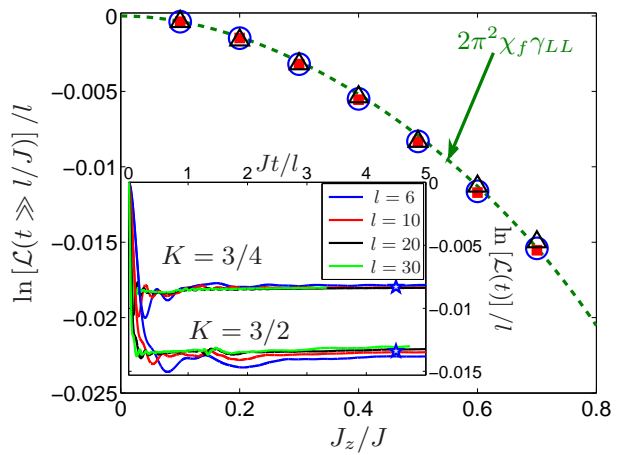}
\caption{The inhomogeneous Loschmidt echo of the XXZ chain for $L=300$ from tDMRG, starting from the XX point to several final $J_z/J$'s for 
$l=10$ (blue circles), 40 (red squares) and 120 (black triangles), the green dashed line is Eq. \eqref{leSQ}. The inset shows 
the time evolution of the LE for $J_z=0.5J$ ($K=3/4$) { and 
$-0.5J$ ($K=3/2$) for several $l$'s, as indicated by the legend. The pentagons indicate the long time asymptotes of $\ln\left[\mathcal L(t\rightarrow\infty )\right]/L$  after a \emph{homogeneous} quench from Ref. \cite{doraLE} in the $L\rightarrow\infty$ limit.}}
\label{leinhxxz}
\end{figure}


\emph{Luttinger model vs. XXZ Heisenberg chain.}
The Luttinger model description of the XXZ Heisenberg chain neglects all sorts of additional terms \cite{giamarchi,nersesyan} (e.g. band curvature or interaction), 
which are always present in lattice models. To establish the reliability of our bosonized calculations,  
we have evaluated the LE after an inhomogeneous quench using tDMRG methods \cite{Vidal2004, Daley2004,FeiguinWhite2005,Schollwock2011}  
for the XXZ Heisenberg chain, and the resulting data is plotted in Fig.~\ref{leinhxxz} for system size $L=300$ and bond dimension $1000$, 
chosen such that finite size and truncation effects are negligible.    
The simulations confirm
the scaling of the echo with the  size of the region $l$ as predicted in Eq.~\eqref{leSQ}. Moreover, even the prefactor of the exponent can be estimated from the fidelity
susceptibility, $\chi_f$, around the XX point of the Heisenberg model, leading to
$q_c/\pi\approx 2\chi_f\pi^2$, where $N$ is the number of lattice
sites and $\chi_f\approx 0.0195$ \cite{sirker}.
This simple estimate describes quantitatively the numerical data
\footnote{Though we have mostly considered even $l$'s on the XXZ chain, let us note that odd 
$l$'s behave slightly differently: the LE decays faster in time before saturation,
though the even-odd difference diminishes with increasing $l$.},
{in spite of the boundary backscattering terms in the inhomogeneous XXZ chain\cite{sedlmayr}, arising from the velocity mismatch between the non-interacting and interacting
regions, regardless to the sign of $J_z$\cite{nersesyan},
as shown in the inset of Fig. \ref{leinhxxz}.}

An  interaction quench over a spatially finite region is realizable experimentally, using a hybrid setup containing a cold atomic LL \cite{cazalillarmp} and a flux qubit proposed in Ref.  \cite{doraLE} to measure the LE after a homogeneous quench.
In addition to an external magnetic field, which tunes the properties of the qubit, the induced magnetic field of the 
flux qubit itself controls the interaction strength in the LL via a Feshbach resonance. 
As we have shown here, by quenching the interaction over a finite spatial region $1/q_c\ll l\ll L$ (and not homogeneously over the whole system), the observation of the 
peculiar scaling of the overlap as in Eq. \eqref{leSQ} in terms of the Luttinger liquid parameters is reachable.

To conclude, we have demonstrated that no orthogonality catastrophe occurs during the time evolution of the Loschmidt echo in Luttinger liquids following an  interaction 
quench within a finite spatial region --- in sharp contrast
to a potential quench. The comparison of the bosonization results  to numerical simulations of the XXZ Heisenberg chain, which contains local backscatterers at the boundary of the quenched region,
 demonstrates the applicability of the Luttinger liquid concept.

\begin{acknowledgments}

This research has been  supported by the Hungarian Scientific  Research Funds Nos. K101244, K105149, K108676, by the ERC Grant Nr. ERC-259374-Sylo and by the Bolyai Program of the HAS.
\end{acknowledgments}

\bibliographystyle{apsrev}
\bibliography{wboson}

\end{document}